\documentclass[12pt]{article}
\usepackage[pdftex]{graphicx}
\usepackage{xspace,colortbl}
\usepackage{epsfig}
\newcommand {\be}{\begin{equation}}
\newcommand {\ee}{\end{equation}}
\newcommand {\ba}{\begin{eqnarray}}
\newcommand {\ea}{\end{eqnarray}}
\date{}
\title{VDM: A model for Vector Dark Matter }
\author{Yasaman Farzan{\footnote{E-mail:
yasaman@theory.ipm.ac.ir}}\,\,\,\,\,and \,\,Amin Rezaei
Akbarieh{\footnote{E-mail: am-rezaei@physics.sharif.ir}}\\{\small
${}^{\star,\dag}$School of physics, Institute for Research in
Fundamental Sciences (IPM)}\\{\small P.O.Box 19395-5531, Tehran,
Iran}\\{\small ${}^\dag$Department of Physics, Sharif University
of Technology}\\{\small P.O.Box 11155-9161, Tehran, Iran}}
\begin{document}
\maketitle
\begin{abstract}
We construct a model based on a new $U(1)_X$  gauge symmetry and a
discrete $Z_2$ symmetry under which the new gauge boson is odd.
The model  contains new complex scalars which carry $U(1)_X$
charge but are singlets of the Standard Model. The $U(1)_X$
symmetry is spontaneously broken but the $Z_2$ symmetry is
maintained, making the new gauge boson a dark matter candidate.
In the minimal version there is only one complex scalar field but by extending the number of scalars to two, the model will enjoy rich
phenomenology which comes in various phases. In one phase, CP is spontaneously broken. In the other phase, an
accidental $Z_2$ symmetry appears which makes one of the scalars stable and therefore a
dark matter candidate along with the vector boson. We discuss the
discovery potential of the model by colliders as well as the
direct dark matter searches.
\end{abstract}
\section{Introduction}
Although in the recent decades overwhelming  evidence for the
presence of dark matter has been accumulated by astrophysical and
cosmological observations, discovering the nature of Dark Matter
(DM) is still one of the open questions before particle
physicists. For example, we still do not know what is the spin of
the DM candidate. Complex and real scalars as well as Dirac and
Majorana fermions as the DM candidates have been extensively
studied in the literature. Vector boson as the dark matter
candidate has only recently received attention
\cite{Hamby,Hambyee,Arina,DiazCruz,Lebedev}. Although the vector boson
playing the role of DM does not necessarily need to be a gauge
vector boson, in most of the scenarios employing a
vector boson as dark matter, it is taken to be the gauge boson of
a new gauge symmetry. In  \cite{Ema} and \cite{ham}, a new
non-Abelian gauge symmetry is introduced in such a way that one \cite{Ema}
or all \cite{ham} of the gauge bosons play the role of the DM.
Abelian gauge boson dark matter has been studied in the context of
extra large dimension \cite{KK}, the little Higgs model
\cite{LittleHiggs} and the linear sigma model \cite{Abe}.

In this paper, we introduce a simple model within which the
Abelian gauge boson plays the role of the dark matter; {\it i.e.}
the gauge group of the standard model is extended to $SU(3)\times
SU(2) \times U(1)_Y \times U(1)_X$ such that all the SM particles are neutral under the $U(1)_X$. The model also contains  new
complex scalar fields which are singlet under the color and
electroweak symmetry but transform under the new $U(1)_X$
symmetry. We will first introduce a model with a single complex scalar and then extend it to two complex scalar bosons. The extension results in a rich phenomenology.  The model also has two $Z_2$ symmetries under which the new
vector boson is odd.  In
other words, the symmetry of the model becomes $U(1)_X\times Z_2
\times Z_2$ times the SM symmetries. One of the new scalar fields
receives a vacuum expectation value which breaks the new
$U(1)_X\times Z_2 \times Z_2$ symmetry into a remnant $Z_2$
symmetry that protects the DM candidate against decay. After the
gauge  symmetry breaking, one of these scalars mixes with the
standard model Higgs. This mixing is the only portal between the
dark sector and the standard model particles.  From the point of
view of direct detection, the standard results for Higgs portal
scenario \cite{ham} applies to this model, too. However, as we
shall see, depending on the parameter range, new annihilation
modes can also be open which do not involve the SM Higgs boson.

 The paper is organized as
follows: In section \ref{model}, we introduce the model and shortly
discuss its various phases. In section \ref{annih}, we discuss the
annihilation of dark matter pair. In section \ref{direct-detection},
we discuss the dark matter direct  and collider searches. In
section \ref{lower}, we discuss the lower bounds on the coupling
with the Standard Model (SM) Higgs. The results are summarized in
section \ref{Con}.
\section{The model\label{model}}
This model is based on a new Abelian  gauge symmetry,  under which
the standard model particles are all neutral. The gauge boson of
the new $U(1)_X$  symmetry is denoted by $V_\mu$. We impose a $Z_2$ symmetry under which the SM particles are all even but $V_\mu$ is odd. As a result, the kinetic mixing between $V_\mu$ and the hyper-charge gauge boson is forbidden by this $Z_2$.
In section \ref{modI}, we introduce the minimal version with a single complex scalar. The possibility is briefly discussed in \cite{Hamby}, too. In section \ref{modII}, we extend the model to include two complex scalars.

\subsection{Minimal model with a single complex scalar\label{modI}} In this model, we include a singlet
scalar $\Phi=(\phi_r+i\phi_i)/\sqrt{2}$ which is charged under $U(1)_X$. The Lagrangian of the scalars and the new vector boson is
\be \mathcal{L}=D_\mu \Phi D^\mu \Phi-\frac{V_{\mu \nu}V^{\mu \nu}}{4}-V(\Phi,H),\ee
where $V_{\mu \nu}=\partial_\mu V_\nu-
\partial_\nu V_\mu$, $D_\mu=\partial_\mu-ig_VV_\mu$ and
\be V=-\mu_\phi^2|\Phi|^2-\mu^2|H|^2+\lambda_\phi|\Phi|^4+\lambda|H|^4+\lambda_{H\phi}|\Phi|^2|H|^2\ .\ee
Notice that the $U(1)_X$ symmetry implies that the Lagrangian is invariant under the $Z_2^{(A)}\times Z_2^{(B)}$ symmetry defined as follows:
$$Z_2^{(A)}: \ \ V_\mu\to -V_\mu \ , \ \ \Phi\to \Phi^* $$
and
$$Z_2^{(B)}: \ \ V_\mu\to -V_\mu \ , \ \ \Phi\to -\Phi^* ,$$
where the rest of the fields are even.

$H$ and $\Phi$ receive VEVs breaking respectively the electroweak and $U(1)_X$ symmetries. Going to the ``unitary'' gauge, the imaginary component of $\Phi$ can be absorbed as the longitudinal component of $V_\mu$. In this gauge, we can write
\ba \Phi=\frac{\phi_r+v_r}{\sqrt{2}} \ {\rm and} \ H=\left( \matrix{ 0 \cr \frac{h+v}{\sqrt{2}}}\right) \ea
where
\begin{eqnarray}
v^2&=&\frac{4\lambda_\phi \mu^2-2\lambda_{H\phi}\mu_\phi^2}{4\lambda\lambda_\phi-\lambda_{H\phi}^2}\cr
v^2_r&=& \frac{4\lambda \mu^2_\phi-2\lambda_{H\phi}\mu^2}{4\lambda\lambda_\phi-\lambda_{H\phi}^2}
\ .\end{eqnarray}
The conditions for successful spontaneous symmetry breaking are $v^2>0$ and $v^{ 2}_r>0$. Notice that while $Z_2^{(B)}$ is broken, the $Z_2^{(A)}$ symmetry still persists making $V_\mu$ a stable particle and therefore a dark matter candidate. The mass of $V_\mu$ is given
 by $$m_V=g_V v_r.$$ $\phi_r$ mixes with $h$ with the following mixing matrix
\ba \frac{1}{2}[ \phi_r \ h]\left[ \matrix{2 \lambda_\phi v^{ 2}_r  & \lambda_{H\phi} vv_r  \cr
\lambda_{H\phi} vv_r  & 2 \lambda v^2}\right] \left[ \matrix{\phi_r
\cr h}\right].\ea
For the case $\lambda_{H\phi}vv_r\ll|2\lambda_\phi v^{ 2}_r-2\lambda v^2|$, the mixing is suppressed and $h$  corresponds
 to $\phi_r$ with mass $\simeq 2\lambda v^2$. The mass of  $\phi_r$ is approximately equal to  $\simeq 2 \lambda_\phi v^{ 2}_r$.

Notice that this model in the minimal version shares some features with a model discussed in \cite{Lebedev} but in \cite{Lebedev} the new scalar degrees of freedom are integrated out. In the present paper, we are more interested in light $\phi_r$.
\subsection{Extended model with two complex scalars \label{modII}}
The model in the extended version
contains two scalar complex fields $\phi_1$ and $\phi_2$ forming a
doublet \begin{eqnarray}\Phi=\left( \matrix{ \phi_1 \cr \phi_2 } \right) \end{eqnarray} which transforms under the
$U(1)_X$ gauge symmetry as
\begin{eqnarray}
\left( \matrix{ \phi_1 \cr \phi_2 } \right) \to U \cdot \left(
\matrix{ \phi_1 \cr \phi_2 } \right)
\end{eqnarray}
where
\begin{eqnarray}
U=\left(\matrix{ \cos \alpha & i \sin \alpha \cr  i \sin \alpha &
\cos \alpha} \right)\ .
\end{eqnarray}

 Notice that under this transformation
$$ \frac{(\phi_1+\phi_2)}{\sqrt{2}} \to e^{i\alpha} \frac{(\phi_1+\phi_2)}{\sqrt{2}}$$
and $$  \frac{(\phi_1-\phi_2)}{\sqrt{2}} \to e^{-i\alpha}
\frac{(\phi_1-\phi_2)}{\sqrt{2}}\ .$$ In other words, in this
model we have two complex scalar fields with opposite $U(1)_X$
charges. However, as we will see, it is more convenient to make
the discussion in terms of $\phi_1$ and $\phi_2$ forming a doublet
representation of $U(1)_X$. In addition to the new $U(1)_X$ gauge
symmetry, we also impose a $Z_2$ symmetry under which all the SM
particles are even and the new particles transform as follows: \be
{Z_2^{(A)}:}\  \Phi \to \sigma_3 \Phi \ {\rm
and} \ \ V_\mu \to - V_\mu \ . \label{Z2A} \ee Out of the doublet
$\Phi$, one can make the following bilinear combinations which all
are invariant under the gauge symmetry:
\begin{equation}
\Phi^\dagger \Phi=\phi_1^\dagger \phi_1+\phi_2^\dagger \phi_2
\label{1} \ee \be \Phi^{T} \sigma_3 \Phi=\phi_1^2-\phi_2^2 \label{3}
\ee and
\begin{equation}
\Phi^\dagger \sigma_1 \Phi=\phi_2^\dagger \phi_1+\phi_1^\dagger
\phi_2 \ .\label{tau1} \ee Notice that $\Phi^{T} \sigma_2 \Phi=0$.
The combinations in Eq.~(\ref{1}) and (\ref{3}) are $Z_2^{(A)}$
even and the one in Eq.~(\ref{tau1}) is $Z_2^{(A)}$ odd. Using
these combinations, the most general potential involving the
scalars of the theory can be written as
\begin{eqnarray}
 V(\Phi,H)=-\mu_H^2 H^\dag H+\lambda_H(H^\dag H)^2-\mu^2\Phi^\dag \Phi
 +\lambda
 (\Phi^\dag
\Phi)^2\label{pot}
\end{eqnarray}
\begin{eqnarray}
&&\hspace{10mm}+\lambda_{H \phi}H^\dag H\Phi^\dag \Phi+\xi^{'}(\Phi^\dag
\sigma_1\Phi)^2\nonumber
\end{eqnarray}
\begin{eqnarray}
+[\xi(\Phi^\dag \Phi)(\Phi^T\sigma_3\Phi)-\mu^{'2}\Phi^T\sigma_3\Phi
+\lambda^{'}(\Phi^T\sigma_3\Phi)^2+\lambda_{H \phi}^\prime H^\dag
H(\Phi^T\sigma_3\Phi)+{\rm h}.c]\nonumber
\end{eqnarray}
where $H$ denotes the  Standard Model Higgs Doublet. Notice
that the mass term of form $\Phi^\dagger \sigma_1 \Phi$ is
forbidden by the $Z_2^{(A)}$ symmetry. The total Lagrangian can be
written as
\begin{eqnarray}
{\cal{L}}={\cal{L}}^{SM}+(D_\mu \Phi)^\dag(D^\mu
\Phi)-V(\Phi,H)-\frac{1}{4}V_{\mu\nu}V^{\mu\nu}
\end{eqnarray} where
the explicit form of covariant derivative is
$D_\mu=\partial_\mu-ig_V {\sigma_1}V_\mu$. The Vector
boson field-strength tensor, $V_{\mu\nu}$, is defined as
$V_{\mu\nu}=\partial_\mu V_\nu-\partial_\nu V_\mu$. Notice that
after imposing the $Z_2^{(A)}$ symmetry as in Eq.~(\ref{Z2A}) ({\it
i.e.,} removing a mass term in the form of  $\Phi^\dagger \sigma_1
\Phi$), there will be an additional $Z_2$ symmetry under which \be
{Z_2^{(B)}:} \ \Phi \to \sigma_3\Phi  \ {\rm
and} \ \ V_\mu \to - V_\mu \ . \label{Z2B} \ee

For simplicity, we take all the couplings to be real. In our
analysis, we make the following conservative assumptions to guarantee the
positiveness of the potential at infinity \be \lambda , \lambda_H
, \lambda_{H \phi} ,\xi^\prime >0 \ \ \ \lambda+2\lambda^\prime>2|\xi|,\
{\rm and} \ \lambda_{H \phi}
>2|\lambda_{H \phi}^\prime|\ . \ee Moreover, we generally assume
$\lambda_{H \phi},\lambda_{H \phi}^\prime \stackrel{<}{\sim} 0.1$ so that the ``SM
Higgs'' still makes sense  within this model.
 Taking $\mu_H^2$, $\mu^2$ and $\mu^{\prime 2}$ (with the
 convention defined in Eq.~\ref{pot}) positive, both $H$ and $\Phi$
 will receive vacuum expectation values breaking  the electroweak
 symmetry as well as the $U(1)_X$ symmetry.
 Using the freedom of $SU(2)\times U(1)$ symmetry, we can go to
 the canonic unitary gauge within which $H^{T}=(0\,\,\,  (v+h)/\sqrt{2})$.
 We can also in general use the global $U(1)_X$ symmetry to absorb
 the imaginary component of $\langle \phi_2\rangle$ and write $\Phi^{T}$ in terms of
 real components as
 \be \label{Phi} \Phi^T=( \frac{ v_{r}+\phi_{r}+iv_{i}+i\phi_{i}}{\sqrt{2}}
 \ \frac{ v^\prime+\phi^\prime_r +i \phi^\prime_i}{\sqrt{2}})\ .\ee
 {A linear combination of these fields will
 be the massless Goldstone boson which can be absorbed as the
 longitudinal component of $V$. }
In this gauge, the new vector boson receives a mass equal to
$g_V\sqrt{ v_{r}^2+v_{i}^2+v^{^\prime 2}}$. The interesting
point is that for a significant part of the parameter space, the
minimum lies at $v^\prime=0$ and $v_{r}^2+v_{i}^2\ne 0$.
In this case, the Goldstone boson is
  $$G\equiv \frac{-v_i \phi'_r+v_r\phi_i'}{\sqrt{v_i^2+v_r^2}}$$
which can be absorbed by using the gauge freedom. The combination
perpendicular to this combination is a mass eigenstate with
nonzero mass:
 $$\phi^\prime\equiv
  \frac{v_r \phi'_r+v_i\phi_i'}{\sqrt{v_i^2+v_r^2}}.$$
By making a local $U(1)_X$ transformation, $G$ can be absorbed and
$\phi_2$ will be of form $\phi^\prime e^{i\beta}$ where
$\beta=\arctan (v_i/v_r)$. As long as $v^\prime=0$, the
$Z_2^{(A)}$ symmetry will be preserved, making the lightest
particle among $\phi^\prime$ and the vector boson stable. Thus,
the vector can contribute to the dark matter content of the
universe if
$$ {g_V^2} (v_r^2+v_i^2)<m_{\phi^\prime}^2.$$
 From now on, we will focus on such a minimum with $v^\prime=0$ and
   we shall assume that $m_V^2<m_{\phi^\prime}^2$.
  Although, the $Z_2^{(A)}$ symmetry is maintained, the $Z_2^{(B)}$
is broken leading to a mixing of the Higgs with $\phi_r$ and/or $\phi_i$
as we will discuss below. Throughout this paper,  we fix one of the mass
eigenvalues $m_{\delta_3}\simeq m_h$ to $125~{\rm GeV}$ as recently announced by the CMS and ATLAS collaborations.


The interaction terms between the gauge boson and scalars are
\begin{eqnarray}
\frac{g_V^2}{2}  V_\mu V^\mu [(\phi_i^2+\phi_r^2+\phi^{\prime
2})+2(\phi_i v_i+\phi_r v_r)]+
\end{eqnarray}
$${g_V} V^\mu \left[-\sin\beta(\phi_r
\partial_\mu\phi^\prime -\phi^\prime
\partial_\mu \phi_r)+\cos \beta (\phi_i
\partial_\mu\phi^\prime -\phi^\prime
\partial_\mu \phi_i)\right].$$
 Depending on the choice of the parameters of the potential,
three phenomenologically  distinct regimes can be realized:
\begin{itemize}
\item \textbf{Phase I} $v^\prime=0$, $v_i,v_r\ne 0$:

 In this case both $\phi_i$ and
$\phi_r$ mix with $h$:
\begin{eqnarray} \label{mixing}
\left(
  \begin{array}{c}
    \phi_r \\
    \phi_i \\
    h\\
  \end{array}
\right)=\left(
          \begin{array}{ccc}
            a_{11} &a_{12} &a_{13} \\
            a_{21} & a_{22} & a_{23} \\
            a_{31} &a_{32}& a_{33}\\
          \end{array}
        \right)\left(
                 \begin{array}{c}
                   \delta_1 \\
                   \delta_2 \\
                   \delta_3 \\
                 \end{array}
               \right)
\end{eqnarray}
where $\delta_i$ are the mass eigenstates.  As a result, both
$\phi_r$ and $\phi_i$ become unstable.
 Notice that in
this case, CP is broken spontaneously. The values of $a_{ij}$ in
terms of the parameters of the potential are given in the
appendix.  We assume that the mixings between the SM Higgs and the
new scalars are small: $a_{31},a_{32},a_{13},a_{23}\ll 1$.

\item \textbf{Phase II}
 $v^\prime=v_r=0$ and $v_i\ne 0$;

 In this case only $\phi_i$ mixes
 with $h$. That is in the matrix shown in Eq. (\ref{mixing}),
 $a_{12}=a_{13}=a_{21}=a_{31}=0$. Taking $\phi_i$ to be CP-even,
 CP will be preserved. For $v_r=0$, the Lagrangian is invariant
 under $\phi_r \to -\phi_r$. Thus, in addition to $Z_2^{(A)}$, there
 is another $Z_2$ which preserves $\delta_1=\phi_r$ against decay.
  As a
 result, there will be two candidates for dark matter:
  $\delta_1=\phi_r$ and
 $V$.
 {In this phase, $\phi^\prime_r$ is the Goldstone boson which can be absorbed
  ({\it i.e.,} $\beta=\pi/2$) so we have $\phi^\prime =\phi^\prime_i$.}
 \item \textbf{Phase III}
 $v^\prime=v_i=0$ and $v_r\ne 0$;

  In this case only $\phi_r$ mixes
 with $h$. That is in the matrix shown in Eq. (\ref{mixing}),
 $a_{12}=a_{23}=a_{21}=a_{32}=0$. CP will be preserved.
 In this phase, $\phi^\prime_i$ is the Goldstone boson
 which can be absorbed
  ({\it i.e.,} $\beta=0$) so we have $\phi^\prime =\phi^\prime_r$.
  It is straightforward to verify that the phases II and III are
  equivalent provided that we substitute
 $$(\mu^2,\lambda_H, \lambda, \xi^\prime,\lambda^\prime, \lambda_{H \phi},\mu^{'2},
\lambda_{H \phi}^\prime,\xi)\to (\mu^2,\lambda_H, \lambda,
\xi^\prime,\lambda^\prime, \lambda_{H \phi},-\mu^{'2},- \lambda_{H \phi}^\prime,-\xi)$$
and $$\phi_i^\prime \leftrightarrow \phi_r^\prime\ .$$
 \end{itemize}

{ In all these phases, in the ``unitary'' gauge where the
Goldstone boson is absorbed, $V_\mu$ is a space-like vector with
three polarizations satisfying $\partial^\mu V_\mu=0$.}

 In the appendix, we have formulated the conditions on the parameters
 of the model under which each of these cases are realized.

\section{Annihilation of dark matter pair \label{annih}}
In the previous section, we introduced the models with one and two scalar fields. In this section,
we discuss how the new stable particles can account for the dark
matter content of the universe for each case one by one.
\subsection{Annihilation modes in the minimal model \label{annI}}
The scalar field, $\phi_r$, can be either lighter or heavier than $V_\mu$. If it is heavier, the main annihilation mode for the $V$ pair will be through $s$-channel scalar exchange. Taking the $\phi_r-h$ mixing small, we find
\begin{eqnarray}
\langle\sigma(V+V\to {\rm
final})v_{rel}\rangle=\frac{64}{3}g_V^4[\frac{\lambda_{H\phi}v v^\prime}{(m_h^2-4m_V^2)(m_{\phi_r}^2-4m_V^2)}]^2F
\end{eqnarray}
with \be F\equiv \lim_{m_{h^*}\rightarrow
2m_V}(\frac{\Gamma(h^*\to final)}{m_{h^*}}).\label{Fdef}\ee Here
$\Gamma(h^*\to final)$ denotes the  rate for the decay mode,
$h^*\rightarrow final$, for a hypothetical SM-like Higgs, $h^*$,
whose mass is $m_{h^*}=2m_V$.
For $m_b< m_{DM}<m_W,m_{\phi_r}$, the main annihilation mode
will be to a $b \bar{b}$ pair \cite{burgess} which is constrained
by bounds on the antiproton flux from PAMELA \cite{Pamela,effe}.
 By adding a new Higgs doublet exclusively coupled to leptons,
the bound can be circumvented but we shall not discuss this
possibility.
  Let us now discuss  the case that
$m_W<m_{V}<m_{\phi_r}$. In this case, the main annihilation mode
is to the $W^+W^-$ pair.
If $m_Z<m_{V}$, the dark matter can also annihilate to a $Z$ pair
with $$\frac{\sigma(V+V\to Z +Z)}{\sigma(V+V\to W^+ +W^-)}= \left.
\frac{ {\rm Br}(H^*\to
 Z
+Z)}{{\rm Br}(H^*\to W^+ +W^-)}\right|_{m_{H^*}=2m_V}<1\ .$$

The subsequent decay of $W^+$ and $W^-$ can produce detectable
secondary particles. In particular, the bounds from the antiproton
and gamma ray fluxes are strong. The present bound from PAMELA on
antiproton flux \cite{Pamela} as well as the gamma ray bound from
Fermi-LAT  \cite{Fermi} lie above 1~pb \cite{Abe}. However, the
forthcoming AMS02 experiment \cite{ams02} may be able to probe
this scenario. Notice that in this range of parameters, the model
shares some features with the model discussed in \cite{Abe} with
the difference that here $g_V$ is a free parameter independent of
the gauge interactions of the standard model so  a wider range of
$m_{DM}$ is possible.

Let us now discuss the case that $m_{\phi_r}<m_{DM}$. In this
case, new annihilation mode for a pair of $V$ become possible:
\be \langle\sigma(V+V\to \phi_r+\phi_r)v_{rel} \rangle= \frac{g_V^4}{24\pi m_V^2} g(m_{\phi_r}^2/m_V^2)\ee
where
$$g(x)=\sqrt{1-x}\left((1+\frac{4}{x-2})^2+\frac{16}{3}\frac{(1-x)^2}{(x-2)^2}+\frac{8}{3}(\frac{1-x}{x-2})(1+\frac{4}{x-2})\right).$$
The produced $\phi_r$ are unstable and will eventually decay to the SM particles.
 If
$2m_b<m_{\phi_r}< 2m_W$, once the $V$ pair annihilate to the
$\phi_r$ pair, the annihilation products will eventually decay
to a $b \bar{b}$ pair leading to an excess in antiproton flux
which is restricted by PAMELA. In the following two ranges, the
antiproton bounds can be avoided: 1) $2m_W<m_{\phi_r}<m_V$; 2)
$m_V\sim$ few GeV and $m_{\phi_r}<2m_p$.
An example of the first range is $$ {\rm Point~ I:}  m_V=250~{\rm GeV}, \ m_{\phi_r}=200~{\rm GeV}, \ v^\prime=1023~{\rm GeV}, \ \lambda_{\phi}=0.13, \ g_V=0.24 $$
and examples of the second range are
$$ {\rm Point~ II:}  m_V=8~{\rm GeV}, \ m_{\phi_r}=1.5~{\rm GeV}, \ v^\prime=187~{\rm GeV}, \ \lambda_{\phi}=0.005 \ g_V=0.042 $$
or
$$ {\rm Point~ III:}  m_V=10~{\rm GeV}, \ m_{\phi_r}=1.5~{\rm GeV}, \ v^\prime=210~{\rm GeV}, \ \lambda_{\phi}=0.005 \ g_V=0.047 $$
where we have taken the mixing given by $\lambda_{H\phi}$ to be small.

\subsection{Annihilation modes in the extended model  \label{annII}}
 As we discussed in
section \ref{modII}, in order for $V_\mu$ to be stable, it should be lighter
than  $\phi^\prime$. However, the other scalars $\delta_i$ (mass
eigenstates made of $\phi_i$, $\phi_r$ and $h$) can be either
lighter or heavier. In case that $\delta_i$ are all heavier than
$V_\mu$, the model can be considered as a Higgs portal model within which
\begin{eqnarray}
\langle\sigma(V+V\to {\rm
final})v_{rel}\rangle=\frac{64}{3}g_V^4[\sum_{j=1}^3\frac{a_{3j}
(a_{1j}v_r+a_{2j}v_i)}{m_{\delta_j}^2-4m_V^2}]^2F
\end{eqnarray}
where $F$ is defined in Eq.~(\ref{Fdef}). The rest of the discussion is as the case with a single scalar.
\\

Let us now discuss the case that $m_{\delta_1}<m_{DM}$. Like the case of a single scalar, in this
case too, new annihilation mode(s) for a pair of $V$ become possible.
\begin{figure}[t]
\vspace{0mm} \hspace{+3.5mm}
\includegraphics[bb=0 0 900 280,scale=0.55]{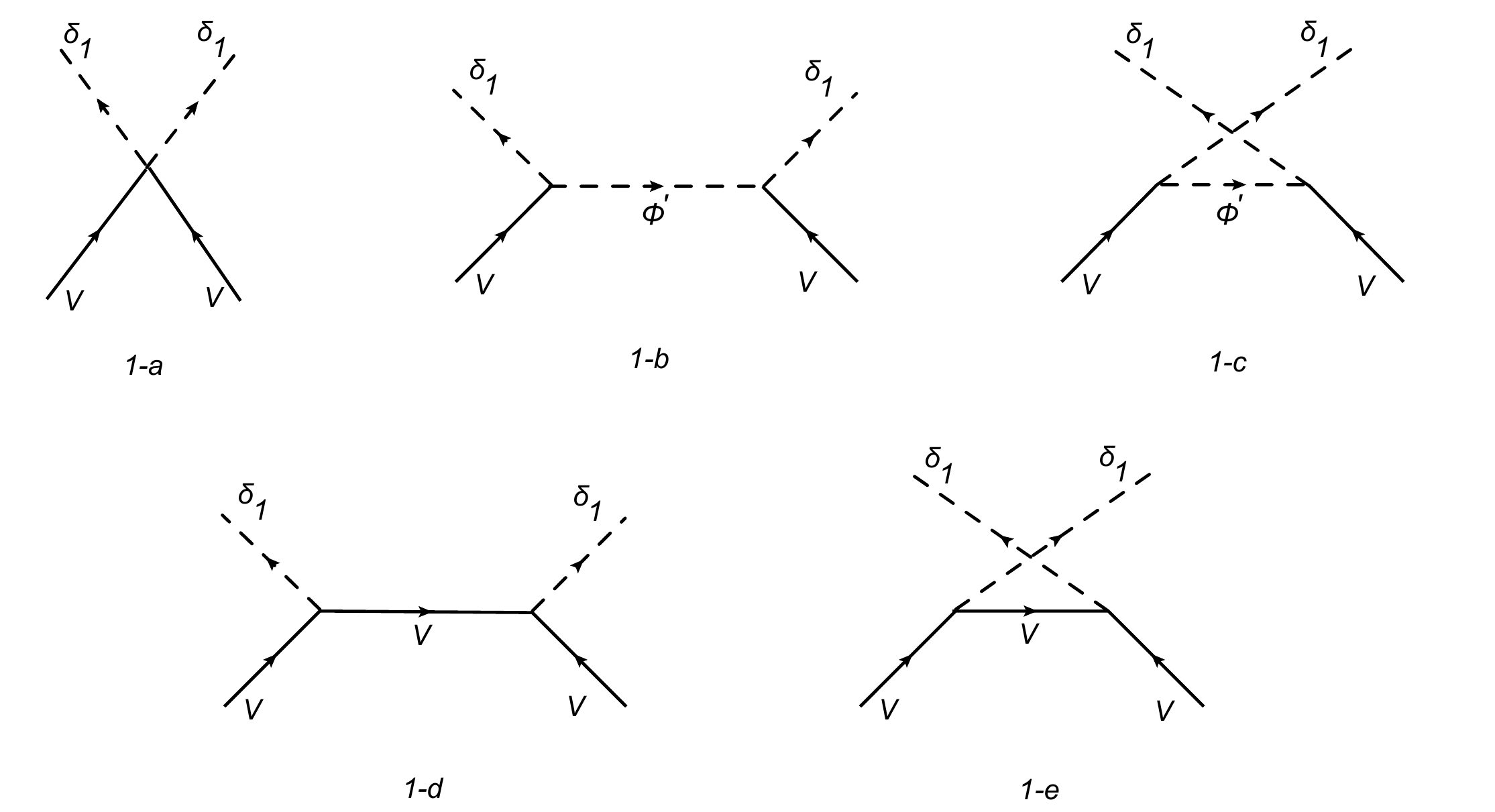}
\caption{Feynman diagrams for the annihilation  of the DM pair to
a $\delta_1$ pair}
\end{figure}\\
The interaction terms within the term $(D_\mu\Phi \cdot D^\mu \Phi)$
lead to the Feynman diagrams shown in Fig.~1. We denote the
amplitude of diagram in Fig.~(1-a) with $M_1$ and the sum of the
amplitudes of diagrams in Figs. (1-b) and (1-c) with $M_2$.
 We denote the sum of the amplitudes of diagrams in Figs (1-d) and (1-e) with $M_3$. There
will be also a contribution from trilinear couplings $A_j
\delta_1^2\delta_j$ which comes from the quartic terms of the
potential in Eq.~(\ref{pot}) as shown in Fig.~2. We denote the
amplitude of diagram in Fig.~2 with $M_4$. The total annihilation cross
section is
\begin{figure}[htb]
\vspace{0mm} \hspace{-25mm}
\includegraphics[bb=0 0 900 280,scale=0.60]{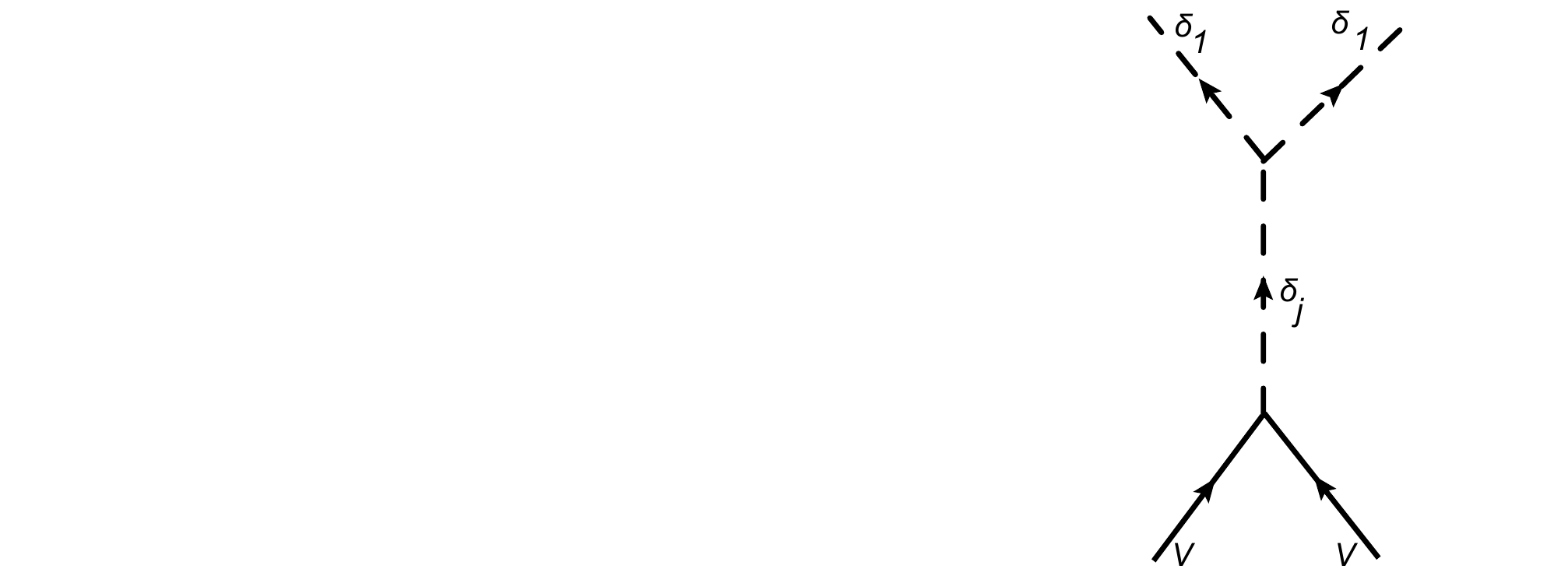}
\caption{The $s$-channel contribution to the DM pair annihilation }
\end{figure}\\
\begin{eqnarray} \label{VVanni}
\langle\sigma_{tot}(V+V\to
\delta_1\delta_1)v_{rel}\rangle&=&\frac{\sqrt{1-\frac{m^2_{\delta_1}}{m^2_V}}}{64\pi
m^2_V}\frac{1}{9}\sum_{spins}|\textit{M}_1+\textit{M}_2+\textit{M}_3+\textit{M}_4|^2\nonumber\\
&=&\sum_{n=1(n\geq m)}^4\sum_{m=1}^4\sigma_{mn}
\end{eqnarray}
where $\sigma_{mn}$ come from the interference of $M_m$ and $M_n$:
\begin{eqnarray}
\sigma_{11}&=&\frac{g_V^4\sqrt{1-\frac{m^2_{\delta_1}}{m^2_V}}}{48\pi
m_V^2}(|a_{11}|^2+|a_{21}|^2)^2\label{sigs}\\
\sigma_{12}&=&\frac{2g_V^4\sqrt{1-\frac{m^2_{\delta_1}}{m^2_V}}}{9\pi
m_V^2}\frac{(m_V^2-m_{\delta_1}^2)S^2(|a_{11}|^2+|a_{21}|^2)}{(m^2_{\phi^{\prime}}+m^2_V-m_{\delta_1}^2)}\nonumber\\
\sigma_{22}&=&\frac{16g_V^4\sqrt{1-\frac{m^2_{\delta_1}}{m^2_V}}}{9\pi
m_V^2}\frac{S^4(m_V^2-m_{\delta_1}^2)^2}{(m^2_{\phi^{\prime}}+m^2_V-m_{\delta_1}^2)^2}\nonumber
\\
\sigma_{13}&=&\frac{g_V^6\sqrt{1-\frac{m^2_{\delta_1}}{m^2_V}}}{18\pi
m_V^4}(|a_{11}|^2+|a_{21}|^2)(a_{11}v_r+a_{21}v_i)^2\frac{m^2_{\delta_1}-4m_V^2}{2m_V^2-m^2_{\delta_1}}\nonumber\\
\sigma_{23}&=&\frac{8g_V^6(1-\frac{m^2_{\delta_1}}{m^2_V})^{3/2}}{9\pi
m_V^2}\frac{S^2(a_{11}v_r+a_{21}v_i)^2}{m^2_{\phi^{\prime}}+m^2_V-m_{\delta_1}^2}\nonumber\\
\sigma_{33}&=&\frac{g_V^8\sqrt{1-\frac{m^2_{\delta_1}}{m^2_V}}}{9\pi
m_V^2}\frac{(a_{11}v_r+a_{21}v_i)^4}{(2m_V^2-m_{\delta_1}^2)^2}[6-4\frac{m_{\delta_1}^2}{m_V^2}+\frac{m_{\delta_1}^4}{m_V^4}]\nonumber\\
\sigma_{14}&=&\frac{g_V^4\sqrt{1-\frac{m^2_{\delta_1}}{m^2_V}}}{12\pi
m_V^2}[\sum_{j=1}^3\frac{A_j(a_{1j}v_r+a_{2j}v_i)}{m^2_{\delta_j}-4m_V^2}](|a_{11}|^2+|a_{21}|^2)\nonumber\\
\sigma_{24}&=&\frac{4S^2g_V^4\sqrt{1-\frac{m^2_{\delta_1}}{m^2_V}}}{9\pi
m_V^2}[\sum_{j=1}^3\frac{A_j(a_{1j}v_r+a_{2j}v_i)}{m^2_{\delta_j}-4m_V^2}]\frac{m^2_V-m^2_{\delta_1}}{m^2_{\phi^{\prime}}+m^2_V-m_{\delta_1}^2}\nonumber\\
\sigma_{34}&=&\frac{g_V^6\sqrt{1-\frac{m^2_{\delta_1}}{m^2_V}}}{9\pi
m_V^4}[\sum_{j=1}^3\frac{A_j(a_{1j}v_r+a_{2j}v_i)}{m^2_{\delta_j}-4m_V^2}]\frac{(a_{11}v_r+a_{21}v_i)^2(4m_V^2-m_{\delta_1}^2)}{2m_V^2-m_{\delta_1}^2}\nonumber\\
\sigma_{44}&=&\frac{g_V^4\sqrt{1-\frac{m^2_{\delta_1}}{m^2_V}}}{12\pi
m_V^2}[\sum_{j=1}^3\frac{A_j(a_{1j}v_r+a_{2j}v_i)}{m^2_{\delta_j}-4m_V^2}]^2,
\nonumber\\
\end{eqnarray}
in which
$S^2=|a_{11}|^2\sin^2\beta+2\Re[a_{11}a_{21}^*]\sin\beta\cos\beta+|a_{21}|^2\cos^2\beta$.
In case $m_{\delta_i}+m_{\delta_k}<2m_V$, the pair of $V$ can
annihilate to $\delta_i+\delta_k$. The annihilation rate will be
given by the same formula with replacement $a_{11}^2\to
a_{1i}a_{1k}$,  $a_{21}^2\to a_{2i}a_{2k}$ and $A_j$ with the
couplings of $\delta_j\delta_k\delta_i$.

Within phase I, all $\delta_i$ are unstable and the discussion is similar to the case of minimal model with only one scalar. However phases II and III will have a totally different phenomenology.
 As we saw earlier, phase III is equivalent
to phase II. In the following, we focus specifically on the
novelties of phase II which applies to phase III, too. Similarly
to phase I, in case that $\delta_1(= \phi_r)$ is lighter than $V$,
a pair of $V$ can annihilate to $\delta_1$. This annihilation mode
will dominate over the annihilation to the SM particles. The cross
section is given by the formulas in Eq.~(\ref{sigs}) setting
$v_r=\cos \beta=0$.
Remember that in this phase $\delta_1(=\phi_r)$ is also stable and
a component of dark matter. The $\delta_1$ pair will annihilate
via Higgs portal with a cross section  \be
\label{IIann}\langle\sigma(\delta_1+\delta_1\to h^* \to{\rm
final})v_{rel}\rangle=\frac{2(\lambda_{H \phi}+2\lambda_{H \phi}^\prime)^2v_H^2}{(4m_{\delta_1}^2-m_h^2)^2}F'\
,\ee where $F'\equiv \lim_{m_{h^*}\rightarrow
2m_{\delta_1}}(\frac{\Gamma(h^*\to final)}{m_{h^*}})$.
 An interesting scenario is the case that
$m_{\delta_1}<m_V<m_{\delta_2},m_{\phi'}$, $$\sigma(V+V\to {\rm
anything})=1~p{\rm b}$$ and $$\sigma( \delta_1+\delta_1\to h^*\to
{\rm final})\gg 1~p{\rm b}.$$ Taking $\lambda_{H \phi}+2\lambda_{H \phi}^\prime\sim
0.1$ this scenario can be realized for {\it e.g.,} the point  in
table 1.
 In this case, at the time of the decoupling of $V$,
$\delta_1$ will be in thermal equilibrium with the SM particles.
Since the annihilation cross section of $\delta_1$ is large, its
density will be suppressed so the dark matter will be mainly
composed of the $V$ particles. The interesting point is that here
even if $2m_b<m_{\delta_1}<2m_W$, the antiproton bound does not
rule out the model as the annihilation of the $V$ pair will lead
to stable $\delta_1$ and the density of $\delta_1$ in the present
time will be too low for its annihilation to lead to a sizeable
secondary flux. For the case that the annihilation cross section
of $\delta_1$ is comparable to that of $V$, both particles will
have comparable densities at the present time. In case that
$m_{\delta_1}>m_V$, the $\delta_1\delta_1$ pair can annihilate to
the $V$ pair. Discussing all these options is beyond the scope of
the present paper and will be done elsewhere.


\begin{table}[htdp]
\caption{Parameters of model. An example for phase II with
$m_{\delta_1}<m_V$.}
\begin{center}
\begin{tabular}{|c|c|c|c|c|c|c|c|c|}
\hline   $\xi$ & $\lambda^\prime$  & $\xi^\prime$ & $\lambda$ &
$\mu$ (GeV)& $\mu^\prime$ (GeV)& $g_V^2$ & $\lambda_{H \phi}$ & $\lambda_{H \phi}^\prime$ \\
\hline
$0.5$ & $0.11$&$0.4$&0.93&409&146&$0.017$ &0.1 &0.1 \\
\hline
\end{tabular}
\end{center}
\label{default}
\end{table}

\begin{table}[htdp]
\caption{The mass spectrum of the point in table 1. All the masses
are in GeV.}
\begin{center}
\begin{tabular}{|c|c|c|c|c|c|}
\hline   $v_i$ & $v_r$  & $m_{\delta_1}$ & $m_{\delta_2}$ &
$m_{\phi'}$ & $m_V$ \\
\hline
 893 &0 & 100& 500&1000 & 116\\
\hline
\end{tabular}
\end{center}
\label{default}
\end{table}

\section{Direct DM detection and signature at the colliders \label{direct-detection}}
In this section, we will discuss the discovery potential of the
model via direct dark matter searches and the production at the
colliders. Since the only portal between the new sector and the
SM particles is through the SM Higgs, both processes are
determined by the couplings $\lambda_{H \phi}$ and $\lambda_{H \phi}^\prime$.
 Let us first discuss the bound from direct searches.
The  interaction of $V$ and the nucleon is through Higgs portal and
is therefore  spin-independent.  Throughout this discussion, we
assume that the vector boson is the main component of the dark
matter. Translating the results of \cite{ham} for vector field in
case of single scalar fermion, we find \be \label{sigmaN} \sigma_N
\equiv\sigma_{SI}(V+ {\rm N} \to V+{\rm N})=\frac{g_V^4M_r^2m_N^2
}{\pi m_V^2 v_H^2}[\frac{\lambda_{H\phi} v
v^{2}_r}{m_h^2m_{\phi_r}^2}]^2f^2\ee where
$M_r=(m_V^{-1}+m_N^{-1})^{-1}$ is the DM-nucleon reduced mass and
$0.14<f<0.66$. Notice that we have taken the mixing between $h$ and
$\phi_r$ small. For two scalar model, we find \be \label{sigmaN}
\sigma_N \equiv\sigma_{SI}(V+ {\rm N} \to V+{\rm
N})=\frac{g_V^4M_r^2m_N^2 }{\pi m_V^2
v_H^2}[(\sum_{j=1}^3\frac{a_{3j}(a_{1j}v_r+a_{2j}v_i)}{m^2_{\delta_j}})]^2f^2\ee
 For the phase II, Eq.~(\ref{sigmaN})
simplifies as
 \be \sigma_{N}=(\lambda_{H \phi}-2\lambda_{H \phi}^\prime)^2\frac{m_V^2M_r^2m_N^2}{\pi
m_{\delta_2}^4m_h^4}f^2.\label{VN}\ee For phase III, the formula
for scattering cross section in Eq.~(\ref{VN}) also applies after
replacing $m_{\delta_2}\to m_{\delta_1}$ and $\lambda_{H \phi}^\prime\to
-\lambda_{H \phi}^\prime$.

In the following, we focus on the model with two scalars. The discussion of the model with a single scalar is very similar to that of phase I if we replace $\delta_1$ with $\phi_r$ and $a_{31}$ with $\lambda_{H\phi}v v_r/(2 \lambda_\phi v^{ 2}_r-2\lambda v^2)$. Of course in the minimal version there is no
$\lambda_{H\phi}^\prime$ coupling and no $\phi^\prime$ or $\delta_2$ field.

Let us now discuss the signature at the high energy colliders. At
the colliders, $\delta_i$ can be in principle produced with a
cross section that is suppressed by $|a_{3i}|^2$ relative to the
SM Higgs of mass $m_{\delta_i}$. For $m_{\delta_i}> 2m_W$ and
relatively large $|a_{3i}|^2$, this may provide an observable
signal. In this case, we expect a
 SM-like Higgs decaying to $W^+W^-$ but with a production rate
 suppressed by a factor of $a_{3i}^2\sim 0.1$. The LHC should be able to put a
 bound on $|a_{3i}|^2$ for this range.
Now, let us discuss another potential discovery channel. If
$\lambda_{H \phi}$ and $\lambda_{H \phi}^\prime$ are relatively large and
some of the new neutral scalar particles are lighter than $m_h/2$,
the Higgs should have sizeable invisible decay modes. For simplicity
let us take $\lambda_{H \phi}^\prime\ll \lambda_{H \phi}$. The invisible decay rate of
the Higgs to first order of approximation will be
$\sim\lambda_{H \phi}^2v_H^2/(64\pi m_h)$. In fact, the bound from the LHC on the Higgs invisible decay rate \cite{grogean} already rules out $\lambda_{H\phi}>0.01$ for $m_{\delta_{1,2}}<m_h/2\simeq$ 63~GeV.

 Let us now discuss each of the  scenarios of interest one
 by one.
\begin{itemize}
\item \textbf{Phase I with  $2m_W<m_{\delta_1}<m_V$}

Within this scenario $\delta_i$ are too heavy to lead to invisible
decay width for the Higgs. However, if $\lambda_{H \phi}, \lambda_{H \phi}^\prime> 0.1$
we might observe two SM-like Higgses decaying to $W^+W^-$ pair at
masses of $m_{\delta_1}$ and $m_{\delta_2}$. For such values of
$\lambda_{H \phi}$ and $\lambda_{H \phi}^\prime$, the scattering cross section off
nuclei is of order of $10^{-46}\lambda_{H \phi}^2(f/0.3)^2~{\rm cm}^2$
which is well below the bound from XENON100 \cite{XENON100}.
\item \textbf{Phase I with  $m_{\delta_1}<m_V$ and $m_{\delta_1}<2m_p$}

At this range, there are light bosons so there is a possibility of
$\delta_3 (\simeq h) \to \delta_1+
\delta_1,\delta_2+\delta_2,\delta_1+\delta_2,
\phi^\prime+\phi^\prime$. The rate depends on the values of
$\lambda_{H \phi}, \lambda_{H \phi}^\prime$. The scattering cross section is expected
to be of order of $10^{-42}\lambda_{H \phi}^2(f/0.3)^2~{\rm cm}^2$ which
is well below the current bounds \cite{XENON100} even for large
$\lambda_{H\phi}$ and $\lambda_{H\phi}^\prime$.
\item \textbf{Phase II with $m_{\delta_1}<m_V$}

Here, $\delta_1$ is stable. If  the scalars are  heavier than
$m_h/2$, the Higgs will not have an invisible decay mode. The
interesting point here is that we require $\lambda_{H \phi}$ and
$\lambda_{H \phi}^\prime$ to be relatively large from constraint on the
$\delta_1$ density.  As result, the model can be tested by
searching for an excess in $W^+W^-$ pair with an invariant mass
equal to $m_{\delta_2}$.
 From the bound on the invisible decay width of the Higgs \cite{grogean}, we already know $m_{\delta_1}>63$ GeV.
 For such values of $\lambda_{H \phi}$ and $\lambda_{H \phi}^\prime$ and for $f=0.3$, the
 scattering cross
 section is about $10^{-45}-10^{-46}$ cm$^2$ which is below the
 current XENON100 bound \cite{XENON100}.
\end{itemize}

\section{Lower bound on coupling of new scalars with Higgs\label{lower}}
Let us first focus on the model with two scalars.
 At the limit $\lambda_{H \phi},
\lambda_{H \phi}^\prime \to 0$, the dark sector including $\Phi$ and $V$ will
decouple from the SM sector. As we saw within phase II,
$\lambda_{H \phi}$ and $\lambda_{H \phi}^\prime$ should be relatively large so that the
lighter dark matter component ($\delta_1$ for the point in table 1
and 2) has sufficiently large annihilation cross section. However,
there is no such a constraint for the minimal model or the phase I of the extended model.
  A lower bound on $\lambda_{H \phi}, \lambda_{H \phi}^\prime$
comes from the assumption that DM is produced thermally in the
early universe through interaction with the Higgs. The production
rate at high temperature is expected to be given by
$O[(\lambda_{H \phi}^2+\lambda_{H \phi}^{\prime ^2})T/(4\pi)]$ which should be compared
with the Hubble expansion rate at the time. Setting the ratio of
the production rate to the Hubble constant at $T=m_{\delta_i}$
larger than one, we find
$$ Max[\lambda_{H \phi},\lambda_{H \phi}^\prime]\stackrel{>}{\sim} 10^{-8}
\left(\frac{m_\delta}{100~{\rm GeV}}\right)^{1/2} \ .$$ The lower
bound  from the decay of the unstable new particles before big
bang nucleosynthesis epoch is much weaker.

The discussion for the minimal model with a single scalar is similar provided that we replace $\delta_i$ with $\phi_r$ and drop $\lambda_{H\phi}^\prime$.

\section{Conclusions\label{Con}}
We have introduced a simple model based on a new $U(1)_X$ gauge
symmetry within which the vector boson plays the role of the dark
matter. The model also contains new complex scalar(s). In the minimal version, there is only one scalar and in the extended version there are two scalars:
$\phi_1$ and $\phi_2$. One of these scalar fields ($\phi_1$ for the extended version)
develops a Vacuum Expectation Value (VEV) which breaks the
$U(1)_X$ symmetry. The dark matter is protected from decay by a
$Z_2$ symmetry. The coupling between the new sector and the SM
sector is through the scalar couplings with the Higgs: $\lambda_{H \phi}$
and $\lambda_{H \phi}^\prime$.

In the extended version, depending on the range of parameters, different phases with
different phenomenology appear. In one phase, both Im($\phi_1$)
and Re($\phi_1$) receive a VEV leading to  spontaneous
CP-violation. In another phase, only  one of Im($\phi_1$) and
Re($\phi_1$) receives a VEV. In the latter case, CP is conserved.
Moreover, a new $Z_2$ appears that leads to a new stable dark
matter candidate. From the abundance of the lighter dark matter
candidate we obtain a lower bound on the couplings of new scalar with the Higgs ({\it i.e.,} on $\lambda_{H \phi}$ and $\lambda_{H \phi}^\prime$)
of order of 0.1.

We have briefly discussed the possibility of discovery at the
colliders and direct dark matter searches. At colliders, the new
particles can show up as invisible decay mode of the SM Higgs or
as extra SM-like  Higgs(es) with a suppressed production rate at
various masses. The scattering cross section off nuclei is
expected to be at least one order of magnitude below the present
bound. Within the phase II of the extended model, the upper bound on the invisible decay width of the Higgs \cite{grogean} implies that the masses of the new particles are larger than $m_h/2=63$~GeV.
\section{Appendix A}
As discussed earlier when $v_i^2+v_r^2\ne 0$, the Higgs field
mixes with $\phi_i$ and/or $\phi_r$ defined in Eq.~(\ref{Phi}).
The components of the mass matrix in the ($\phi_r \ \phi_i \ h$) basis
are \begin{eqnarray} M_{11}^2 &=&-\mu^2-2\mu^{'2}+(\lambda -6
\lambda^\prime) v_i^2+
3(\lambda+2\lambda^\prime+2\xi)v_r^2+(\lambda_{H \phi}+2\lambda_{H \phi}^\prime)\frac{v_H^2}{2}
\cr
 M_{22}^2 &=&-\mu^2+2\mu^{'2}+3(\lambda +2
\lambda^\prime -2\xi)v_i^2+
(\lambda-6\lambda^\prime)v_r^2+(\lambda_{H \phi}-2\lambda_{H \phi}^\prime)\frac{v_H^2}{2}\cr
 M_{33}^2 &=&-\mu_H^2+3\lambda_H
 v_H^2+\frac{\lambda_{H \phi}}{2}(v_i^2+v_r^2)+\lambda_{H \phi}^\prime(v_r^2-v_i^2)\cr
 M_{12}^2 &=& 2(\lambda-6\lambda^\prime)v_iv_r \cr
M_{13}^2 &=& (\lambda_{H \phi}+2\lambda_{H \phi}^\prime)v_Hv_r \cr M_{23}^2 &=&
(\lambda_{H \phi}-2\lambda_{H \phi}^\prime)v_Hv_i\ . \cr\nonumber
\end{eqnarray}
The components of mass matrix of $(\phi'_r,\phi'_i)$ are
\begin{eqnarray}
\label{primemass}M_{rr}^2&=&-\mu^2+2\mu^{'2}+
v_i^2(\lambda+2\lambda'-2\xi)+v_r^2(\lambda-2\lambda'+2\xi')
+\frac{v_H^2}{2}(\lambda_{H \phi}-2\lambda_{H \phi}^\prime) \cr
M_{ii}^2&=&-\mu^2-2\mu^{'2}+
v_i^2(\lambda-2\lambda'+2\xi^\prime)+v_r^2(\lambda+2\lambda'+2\xi)
+\frac{v_H^2}{2}(\lambda_{H \phi}+2\lambda_{H \phi}^\prime)\cr
M_{ir}^2&=&2(\xi^\prime+2\lambda^\prime)v_iv_r
 . \end{eqnarray}

In the limit $\lambda_{H \phi},\lambda_{H \phi}^\prime \ll 1$, we can write
$m_h^2\simeq M_{33}^2$ and the mixing matrix in Eq.~(\ref{mixing})
approximately as
 \begin{eqnarray}
 \left[ \matrix{ \cos \theta & \sin \theta & a_{13} \cr
 -\sin \theta & \cos \theta & a_{23} \cr
a_{31} & a_{32} & 1} \right]
\end{eqnarray}
where
$$\tan 2 \theta=\frac{2M_{12}^2}{M_{22}^2-M_{11}^2}$$
and  \begin{eqnarray} a_{13}&=&\frac{(\lambda_{H \phi}
S_1+2\lambda_{H \phi}^\prime S_2)v_H}{(m_h^2-m_{\delta_1}^2)(m_h^2-m_{\delta_2}^2)}
\cr a_{23}&=&\frac{(\lambda_{H \phi}T_1+2\lambda_{H \phi}^\prime T_2)v_H}{
(m_h^2-m_{\delta_1}^2)(m_h^2-m_{\delta_2}^2) }
\end{eqnarray}
in which
 \begin{eqnarray}
 S_1&=&-M_{33}^2(\cos\theta v_r+\sin \theta
 v_i)+M_{22}^2v_r\cos\theta+M_{11}^2v_i
 \sin\theta-M_{12}^2(v_r\sin\theta+v_i \cos\theta)\cr
 S_2&=&-M_{33}^2(\cos\theta v_r-\sin \theta
 v_i)+M_{22}^2v_r\cos\theta-M_{11}^2v_i
 \sin\theta-M_{12}^2(v_r\sin\theta-v_i \cos\theta)\cr
  T_1&=&M_{33}^2(\sin\theta v_r-\cos \theta
 v_i)-M_{22}^2v_r\sin\theta+M_{11}^2v_i
 \cos\theta-M_{12}^2(v_r\cos\theta-v_i \sin\theta)\cr
 T_2&=&M_{33}^2(\sin\theta v_r+\cos \theta
 v_i)-M_{22}^2v_r\sin\theta-M_{11}^2v_i
 \cos\theta-M_{12}^2(v_r\cos\theta+v_i \sin\theta).\cr
\end{eqnarray}
Finally,
\begin{eqnarray}
a_{31}&=&a_{23}\sin\theta-a_{13}\cos\theta\cr
a_{32}&=&-a_{23}\cos\theta-a_{13}\sin\theta\ .
\end{eqnarray}
In phase I, we have
$$ v_r^2=\frac{\mu^2(4\lambda^\prime-\xi)+2\mu^{\prime 2}
(\lambda-2\lambda^{\prime}-2\xi)}{2(4\lambda \lambda'-8\lambda'^2-\xi^2)},
$$
$$v_i^2=\frac{\mu^2(4\lambda^\prime+\xi)+2\mu^{\prime2}
(-\lambda+2\lambda^{\prime}-\xi)}{2(4\lambda \lambda'-8\lambda'^2-\xi^2)},
$$
and \be m_{\phi'}^2=
\frac{4(2\lambda^\prime+\xi^\prime)(2\lambda^\prime \mu^2-\xi
\mu^{\prime 2})} {4\lambda \lambda'-8\lambda'^2-\xi^2}.\ee Of
course, in order for the phase I to be realized, the parameters
have to be in a range where $v_i^2,v_r^2,m_{\phi'}^2>0$. The
stability of $V$ ({\it i.e.,}
$m_V^2=g_V^2(v_i^2+v_r^2)<m_{\phi'}^2$) then implies \be
g_V<\sqrt{2(2\lambda'+\xi')}\ .\ee\\

 In order for the phase II with $v_r=v'=0$ to be realized,
 we should have
 $$\mu^2-2\mu^{'2}>0 \ , \
 \mu^2(\xi+\xi^\prime-2\lambda^\prime)+2\mu^{'2}(\xi+\xi^\prime-\lambda)>0$$
 and
 $$\mu^2(\xi-4\lambda^\prime)+2\mu^{'2}(\xi+2\lambda^\prime-\lambda)>0\ .$$

 In phase II, we find
$m_{\delta_1}^2=M_{11}^2$, $m_{\delta_2}^2\simeq M_{22}^2$,
$m_h^2\simeq m_{\delta_3}^2\simeq M_{33}^2$ and
$a_{12}=a_{21}=a_{13}=a_{31}=0$ and
$$a_{23}=-a_{32}=\frac{(\lambda_{H \phi}-2\lambda_{H \phi}^\prime)v_iv_H}{m_{\delta_2}^2-m_h^2}\ .$$

Within the phase II
$$v_i^2=\frac{\mu^2-2\mu'^2}{\lambda+2\lambda'-2\xi}$$
and \be m_{\phi'}^2=\frac{\mu^2(2 \xi+2\xi'-4\lambda')
+\mu^{'2}(4\xi-4\lambda-4\xi^\prime)}{\lambda+2\lambda^\prime-2\xi}.\ee
The stability of $V$ ({\it i.e.,}
$m_V^2=g_V^2(v_i^2+v_r^2)<m_{\phi'}^2$) then implies \be
g_V<\left(\frac{2[\mu^2(\xi+\xi^\prime-2\lambda^\prime)
+\mu'^2(2\xi-2\xi'-2\lambda)]}{\mu^2-2\mu'^2}\right)^{1/2} .\ee

In phase III where $v_r\ne 0$ and $v_i= 0$, we find
$m_{\delta_2}^2=M_{22}^2$, $m_{\delta_1}^2\simeq M_{11}^2$,
$m_h^2\simeq m_{\delta_3}^2\simeq M_{33}^2$ and
$a_{12}=a_{21}=a_{23}=a_{32}=0$ and
$$a_{13}=-a_{31}=\frac{(\lambda_{H \phi}+2\lambda_{H \phi}^\prime2)v_rv_H}{m_{\delta_1}^2-m_h^2}\
.$$\\
Within this phase
\begin{eqnarray}
v_r^2=\frac{\mu^2+2\mu^{'2}}{\lambda+2(\lambda^\prime+\xi)}
\end{eqnarray}
and
\begin{eqnarray}
m^2_{\phi^{'}}=\frac{2\mu^2(\xi^{'}-\xi-2\lambda^{'})+4\mu^{'2}(\xi+\xi^{'}+\lambda)}
{\lambda+2(\lambda^\prime+\xi)}.
\end{eqnarray}
The stability of $V$ ({\it i.e.,}
$m_V^2=g_V^2(v_i^2+v_r^2)<m_{\phi'}^2$) then implies
\begin{eqnarray}
g_V<(\frac{2[\mu^2(\xi^{'}-\xi-2\lambda^{'})+2\mu^{'2}(\lambda+\xi+\xi^{'})]}{\mu^2+2\mu^{'2}})^{1/2}.
\end{eqnarray}
\section*{Acknowledgment}
Y.F. acknowledges partial support from the  European Union FP7 ITN
INVISIBLES (Marie Curie Actions, PITN- GA-2011- 289442) and thanks Galileo Galilei Institute for Theoretical Physics for its hospitality.
She is grateful to ICTP for the partial financial support and the hospitality of its staff.
She is also grateful to Prof. Hambye and Prof. de Gouvea for useful comments. A. R. is
grateful to Prof. M. Golshani for  encouragement and support.

\end{document}